\documentclass[12pt]{iopart}

\usepackage{graphicx}
\usepackage{dcolumn}
\usepackage{bm}

\begin{document}

\title[Dissipation of superfluid flow in a periodically-dressed BEC]{
Anisotropic dissipation of superfluid flow in a periodically-dressed Bose-Einstein condensate}

\author{D.M. Stamper-Kurn}

\address{Department of Physics, University of California, Berkeley
CA 94720}

\date{\today}

\begin{abstract}
The introduction of a steady-state spatially-periodic Raman
coupling between two components of an ultracold atomic gas
produces a dressed-state gas with an anisotropic and tunable
dispersion relation.  A Bose-Einstein condensate formed in such a
gas is consequently characterized by an anisotropic superfluid
critical velocity.  The anisotropic dissipation of superfluid
flow is quantified by considering the scattering of impurities
flowing through this superfluid.  A gradual transition from the
isotropic nature of an uncoupled Bose-Einstein condensate to the
anisotropic periodically-dressed condensate is obtained as the
strength of the Raman coupling is varied.  These results present
a clear signature for future experiemental realizations of this
novel superfluid.
\end{abstract}

\pacs{03.75.Fi, 05.30.Jp}

 \maketitle

The experimental attainment of quantum degenerate dilute gases,
composed of both bosonic and fermionic atoms, has created many
new opportunities in the study of quantum fluids.  In particular,
recent years have seen a flurry of experimental studies of
superfluidity in dilute, scalar Bose-Einstein condensates.  These
include the observations of critical velocities for superfluid
flow about microscopic \cite{chik00} and macroscopic \cite{rama99}
obstacles, the onset of turbulent flow above this critical
velocity \cite{onof00},  beautiful studies of quantized vortices
and vortex lattices in rotating Bose-Einstein condensates
\cite{matt99vort,madi99vort,abos01lattice}, and other
manifestations of irrotational superfluid flow
\cite{mara00scissors,hech02irrot}.  These experiments  are
closely analogous to those performed on another scalar
superfluid, liquid $^4$He, but apply new experimental probes and
allow new insights due to the vastly different parameter regime of
the dilute atomic gases, and their amenability to a new set of
tools for manipulation and probing.

Dilute atomic gases also offer the opportunity to create novel
quantum fluids by using these various tools to manipulate the
internal and external states of the atoms comprising the
superfluid.  For example, different types of multi-component
condensates have been studied: externally-coupled two-component
condensates of $^{87}$Rb at JILA (reviewed in \cite{corn99var}),
and $F=1$ spinor condensates of sodium at MIT (reviewed in
\cite{stam00leshouches}).  One experiment at JILA explored how a
spatially-selective coupling between two trapped hyperfine levels
of a rubidium condensate leads to a ``winding'' of the order
parameter, akin to the phase winding of the order parameter which
occurs due to rotation about a vortex core, and how this winding
is recurrently ``undone'' due to the motional dynamics of a
two-component condensate \cite{matt99twist}.  This observation
illustrates that vortices in such a novel system should not be
metastable (similarly predicted for spinor condensates
\cite{ho98}), a major modification from the superfluid behaviour
of a single component gas.

In a similar vein, we have previously considered the novel
superfluid properties introduced to a two-component Bose
condensed atomic gas which is coupled by a spatially periodic
coupling field \cite{higb02}.  This spatially-periodic coupling
results from intersecting, non-collinear laser beams which induce
a Raman coupling between two internal states of the ultracold
gas.  The Raman coupling introduces an anisotropy to this
periodically-dressed fluid which should manifest itself in an
anisotropic superfluid critical velocity.  Controlling various
parameters of the Raman coupling laser beams allows one to
dramatically alter this critical velocity, leading to the
dissipation of superfluid flow at velocities much lower than the
critical velocity for the uncoupled Bose condensate.

In this paper, we develop further a theory describing the
dissipation of superfluid flow in a periodically-dressed
Bose-Einstein condensate.  Using the Bogolibov approximation
theory which was introduced in Ref.\ \cite{higb02}, the possible
disspation of superfluid flow is treated by analyzing the
scattering of massive impurities which flow through the
superfluid.  In analogy with similar calculations for a scalar
Bose-Einstein condensate \cite{timm97symp,chik00,kett00coll},
impurity scattering at low impurity velocities is found to be
suppressed by the energetics of the quasi-particle spectrum
(related to the Landau criterion for superfluidity
\cite{land41}), by the suppression of density fluctuations in the
low-momentum phonon regime \cite{nozi90}, and by variations of
the internal-state compositions of quasiparticles at different
wavevectors.  These effects combine to give a smooth evolution in
the disspation rates for superfluid flow as the Raman coupling
strength is increased, gradually converting an uncoupled two
component Bose-Einstein condensate (essentially a scalar
superfluid) to a periodically-dressed condensate with
dramatically different superfluid properties.  This superfluid
dissipation rate gives an appealingly clear signature for future
experimental studies.

\section{Quasi-particle modes of the periodically-dressed condensate}

\begin{figure}[htbf]
\begin{center}
\includegraphics[width=3in]{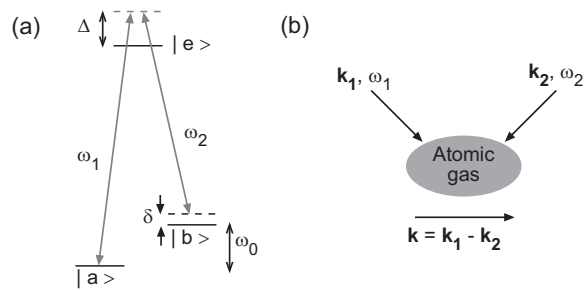}
\end{center}
     \caption{Engineering propeties of a
     periodically-dressed atomic gas.  (a) Laser beams of
     frequency $\omega_1$ and $\omega_2$ induce Raman
     transitions between internal states $|a\rangle$ and
     $|b\rangle$.  The beams share a common large detuning $\Delta$ from the excited atomic state,
     and this state is adiabatically eliminated in the theoretical treatment.
     The lasers are detuned by an amount $\delta$ from the two-photon Raman transition.
     (b) Such a Raman transition imparts a momentum transfer of
     $\hbar {\bf{k}} = \hbar ({\bf{k}}_1 - {\bf{k}}_2)$, where ${\bf{k}}_1$ and
     ${\bf{k}}_2$ are the wavevectors of the Raman coupling
     lasers.
     }
     \label{fig:scheme}
\end{figure}

As presented in Ref.\ \cite{higb02}, a periodically-dressed
Bose-Einstein condensate (PDBEC) is formed of atoms of mass $m$
with two internal states, $A$ and $B$, separated by an energy
difference $\hbar \omega_0$.  These atoms are exposed
continuously to two laser beams of wavevectors ${\bf{k}}_1$ and
${\bf{k}}_2$ and frequencies $\omega_1$ and $\omega_2$, which
couple the states $A$ and $B$ via a Raman transition (see Figure
\ref{fig:scheme}).  In such a Raman transition, an atom in
internal state $A$ with wavevector ${\bf{q}} - {\bf{k}}/2$
absorbs a photon from beam $1$ and emits a photon into beam $2$,
arriving in internal state $B$ with wavevector ${\bf{q}} +
{\bf{k}}/2$ gaining a momentum $\hbar {\bf{k}} = \hbar
({\bf{k}}_1 - {\bf{k}}_2)$ and a kinetic energy $\hbar \delta =
\hbar (\omega_1 - \omega_2 - \omega_0)$.

Keeping in mind this coupling between atoms of different momenta,
we let $a_{\bf{q}}$ ($a_{\bf{q}}^\dagger$) be the creation
(annihilation) operator for an atom in internal state $A$ and
wavevector ${\bf{q}} - {\bf{k}}/2$, and similarly let $b_{\bf{q}}$
($b_{\bf{q}}^\dagger$) be the creation (annihilation) operator for
an atom in internal state $B$ and wavevector ${\bf{q}} +
{\bf{k}}/2$.  We refer to these as the bare-state creation and
annihilation operators.  The justification for this particular
notation lies in the dressed-state picture \cite{cohe92}, in
which one considers the driving optical fields to be part of the
quantum system.  In this picture, the Raman transition connects
the states $|a_{\bf{q}}\rangle  = |A, {\bf{q}} - {\bf{k}}/2;
N_1+1, N_2\rangle$ and $|b_{\bf{q}}  = |B, {\bf{q}} + {\bf{k}}/2;
N_1, N_2+1\rangle$, where $|a_{\bf{q}}\rangle$ represents an atom
in state $A$ and wavevector ${\bf{q}} - {\bf{k}}/2$ and an optical
field with $N_1+1$ photons in beam $1$ and $N_2$ photons in beam
$2$, and similarly for $|b_{\bf{q}}\rangle$.  In the
dressed-state picture, $|a_{\bf{q}}\rangle $ and
$|b_{\bf{q}}\rangle$ are states for which the total momentum of
the atom + photon system is equal -- by a proper choice of
intertial frame, this total momentum is taken to be  $\hbar
{\bf{q}}$).  The states are separated by an energy difference
$\hbar \delta$, and coupled by a static and (in the dressed-state
picture) spatially uniform Raman coupling.  The system is then
represented by a many-body Hamiltonian of the form
\begin{equation}
\label{eq:manybodyhami} {\mathcal{H}} = \sum_{\bf{q}} \left[
\left( \frac{\hbar^2}{2 m} \left({\bf{q}} -
\frac{{\bf{k}}}{2}\right)^2 + \frac{\hbar \delta}{2} \right)
a^\dagger_{\bf{q}} a_{\bf{q}} + \left(\frac{\hbar^2}{2 m}
\left({\bf{q}} + \frac{{\bf{k}}}{2}\right)^2 - \frac{\hbar
\delta}{2} \right) b^\dagger_{\bf{q}} b_{\bf{q}} + \frac{\hbar
\Omega}{2} \left( b^\dagger_{\bf{q}} a_{\bf{q}} +
a^\dagger_{\bf{q}} b_{\bf{q}}\right) \right]
\end{equation}
Here $\Omega$ is the two-photon Rabi frequency characterizing the
strength of the Raman transition, determined by the polarizations
and intensities of the driving laser beams and by dipole matrix
elements of the atoms.

This Hamiltonian is diagonalized by transforming to the
dressed-state creation and annihilation operators $\mu_{\bf{q}}$,
$\mu^\dagger_{\bf{q}}$,
 $\pi_{\bf{q}}$, $\pi^\dagger_{\bf{q}}$ which are defined as
 linear combinations of the bare-state operators
\begin{equation}
\left( \begin{array}{c} a_{\bf{q}}\\
b_{\bf{q}}\end{array} \right) =
\left( \begin{array}{c c} \cos \theta_q/2 & \sin \theta_q/2 \\
-\sin \theta_q/2 & \cos \theta_q/2 \end{array} \right) \left( \begin{array}{c} \mu_{\bf{q}} \\
\pi_{\bf{q}} \\
 \end{array} \right)
\label{eq:rotation}
\end{equation}
The mixing angle $\theta_q$ is defined by the relation $\tan
\theta_q = \omega / (\delta - \hbar {\bf{q}} \cdot {\bf{k}} /
m)$.  The dressed states have energies
\begin{equation}
\hbar \omega_{\bf{q}}^\pm = \frac{\hbar^2}{2 m} \left( q^2 +
\frac{k^2}{4} \right) \pm \frac{\hbar}{2} \sqrt{\left( \delta -
\frac{\hbar {\bf{q}} \cdot {\bf{k}}}{m} \right)^2 + \Omega^2}
\end{equation}
Here, the minus sign refers to the lower dressed-state dispersion
curve and to the operators $\mu_{\bf{q}}$ and
$\mu^\dagger_{\bf{q}}$, while the plus sign refers to the upper
dressed-state dispersion curve and to the operators
$\pi_{\bf{q}}$ and $\pi^\dagger_{\bf{q}}$.

A Bose-Einstein condensate formed from this periodically-dressed
atomic gas has a macroscopic population of $N_0$ atoms in the
lowest energy momentum state: this state lies on the lower
dressed-state dispersion curve and is taken to have momentum
$\hbar {\bf{Q}}$ \cite{degeneracyfootnote}.  In Ref.\
\cite{higb02}, a Bogoliubov-approximation theory was developed to
account for the effects of weak interatomic interactions on a
PDBEC. An interaction Hamiltonian of the form
${\mathcal{H}}_{\mbox{int}} = \frac{g}{2} \sum_{{\bf{q}}}
(n_{\bf{q}} n_{-{\bf{q}}} - N)$ was considered, where $N$ is the
total number of atoms in the system and $n_{\bf{q}} =
\sum_{\bf{k}} \left( a^\dagger_{{\bf{k}} + {\bf{q}}}
 a_{{\bf{k}}}
+  b^\dagger_{{\bf{k}} + {\bf{q}}}
 b_{{\bf{k}}}\right)$ is the Fourier transform of the
 density operator.
Quasi-particle energies and operators were found by diagonalizing
a $4 \times 4$ matrix $(H_{\bf{q}})_{i j} =
({\mathcal{E}}_{\bf{q}})_{i j} + \mu x_i x_j$ where
${\mathcal{E}}_{\bf{q}}$ is a diagonal matrix with entries
$({\mathcal{E}}_{\bf{q}}^-, -{\mathcal{E}}_{-\bf{q}}^-,
{\mathcal{E}}_{\bf{q}}^+, -{\mathcal{E}}_{-\bf{q}}^+)$ with
${\mathcal{E}}^\pm_{\bf{q}} = \hbar (\omega_{{\bf{Q}} +
\bf{q}}^\pm - \omega_{\bf{Q}}^-)$, $\mu = g N$ is the chemical
potential, and $x = (\cos \Delta_{\bf{q}}, -i \cos
\Delta_{-{\bf{q}}} , \sin \Delta_{\bf{q}}, -i \sin
\Delta_{-{\bf{q}}} )$, where $\Delta_{\bf{q}} = (
\theta_{{\bf{Q}}+{\bf{q}}} - \theta_{\bf{Q}})/2$. Diagonalization
yields
\begin{equation}
H_{\bf{q}} = M_{\bf{q}} \tilde{H}_{\bf{q}} M_{\bf{q}}^{-1}
\end{equation}
where $\tilde{H}_{\bf{q}}$ is a diagonal matrix with entries
$\hbar (\tilde{\omega}_{{\bf{Q}} + \bf{q}}^-,
-\tilde{\omega}_{{\bf{Q}}-\bf{q}}^-, \tilde{\omega}_{{\bf{Q}}+
\bf{q}}^+, -\tilde{\omega}_{{\bf{Q}}-\bf{q}}^+)$ by which the
lower (minus sign) and upper (plus sign) quasi-particle energies
$\hbar \tilde{\omega}_{{\bf{Q}}+ \bf{q}}^{\pm}$ are defined.  One
also obtains the creation and annihilation operators for the lower
($\tilde{\mu}_{\bf{q}}$, $\tilde{\mu}_{\bf{q}}^\dagger$) and upper
($\tilde{\pi}_{\bf{q}}$, $\tilde{\pi}_{\bf{q}}^\dagger$) branch
quasi-particles through the following relation:
\begin{equation}
\left( \begin{array}{c} a_{{\bf{Q}} + {\bf{q}}} \\
i a_{{\bf{Q}} - {\bf{q}}}^\dagger \\
b_{{\bf{Q}} + {\bf{q}}} \\
i b_{{\bf{Q}} - {\bf{q}}}^\dagger \end{array} \right) =
R_{\bf{q}} \left( \begin{array}{c} \mu_{{\bf{Q}} + {\bf{q}}} \\
i \mu_{{\bf{Q}} - {\bf{q}}}^\dagger \\
\pi_{{\bf{Q}} + {\bf{q}}} \\
i \pi_{{\bf{Q}} - {\bf{q}}}^\dagger \end{array} \right) =
R_{\bf{q}} M_{\bf{q}} \left( \begin{array}{c} \tilde{\mu}_{{\bf{Q}} + {\bf{q}}} \\
i \tilde{\mu}_{{\bf{Q}} - {\bf{q}}}^\dagger \\
\tilde{\pi}_{{\bf{Q}} + {\bf{q}}} \\
i \tilde{\pi}_{{\bf{Q}} - {\bf{q}}}^\dagger \end{array} \right)
\label{eq:baretoquasi}
\end{equation}
Here $R_{\bf{q}}$ contains
trigonometric functions of $\theta_{{\bf{Q}}+{\bf{q}}}/2$ and
$\theta_{{\bf{Q}} - {\bf{q}}}/2$ in accordance with Eq.\
\ref{eq:rotation}.  The Bose commutation relations take the form
\begin{equation}
\left[ \chi_i, \chi_j^\dagger \right] = A_{i j}  = \left\{
\begin{array}{c l} 0 & i \neq j \\
1 & i = j = 1 \mbox{ or } i = j = 3 \\
-1 & i = j = 2 \mbox{ or } i = j = 4 \end{array} \right.
\end{equation}
where $\chi$ represents any of the four-component vectors in Eq.\
\ref{eq:baretoquasi}.  These relations are clearly preserved by
the rotation matrix, i.e.\ $R_{\bf{q}} A R_{\bf{q}}^\dagger = A$,
while the imposition of Bose commutation relations for the
quasi-particle operators enforces a normalization criterion
$M_{\bf{q}} A M_{\bf{q}}^\dagger = A$.

Experimentally, values for the Raman detuning $\delta$, the Rabi
frequency $\Omega$, the chemical potential $\mu$ and the Raman
momentum transfer $\hbar k$ can be chosen over a wide range of
parameters.  For example, let us consider a particular
realization of a PDBEC using two hyperfine states of $^{87}$Rb.
One may consider the case of Raman excitation using two lasers
both tuned near the $D2$ optical transition at a wavelength
$\lambda \simeq 780$ nm.  Thus, the Raman recoil energy $E_k =
\hbar^2 k^2 / 2 m$ can be readily tuned in the range $0 \leq E_k
\leq h \times 15$ kHz.  Properties of a PDBEC are governed by the
ratios $\hbar \delta / E_k$, $\Omega / E_k$ and $\mu / E_k$. The
Raman detuning $\delta$ and the Rabi frequency $\Omega$ can be
chosen nearly arbitrarily, reaching maximum values which are
orders of magnitude larger than $E_k$.  The chemical potential
$\mu = 4 \pi \hbar^2 a n / m$ has a typical value of 1 kHz for a
density of $n \simeq 1 \times 10^{14} \, \mbox{cm}^{-3}$ and a
scattering length $a \simeq 100$ bohr -- thus the ratio $\mu/E_k$
can be as small as $\sim 1/10$ for counter-propagating Raman
excitation lasers, and greatly increased by using Raman beams
intersecting at small angles.

\begin{figure}[htbf]
\begin{center}
\includegraphics[width=3in]{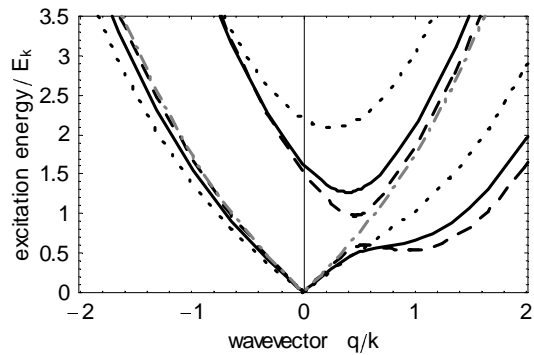}
\end{center}
     \caption{Quasiparticle dispersion relations for a PDBEC.
     Shown are the energies of quasiparticle in the lower (created by $\mu_{{\bf{Q}} + {\bf{q}}}$) and
     upper (created by $\pi_{{\bf{Q}} + {\bf{q}}}$) dispersion
     curves with momentum $\bf{q}$ with respect to the condensate.
      Excitations parallel to the Raman momentum transfer are
      considered, with the definition $q = {\bf{q}} \cdot {\bf{k}} / k$.  Black
      curves represent PDBEC's with $\hbar \delta / E_k = 1/2$,
      $\mu = 1$ and $\hbar \Omega / E_k = 1/2$ (dashed), 1 (dotted)
      and 2 (solid).  The gray dot-dashed curve shows the Bogoliubov
      excitation spectrum for a single component condensate with
      $\mu = 1$.
     }
     \label{fig:quasidisprelations}
\end{figure}

For illustration, we show in Figure \ref{fig:quasidisprelations}
the quasiparticle dispersion relations for a PDBEC with a
chemical potential $\mu/E_k = 1$, a Raman detuning of of $\hbar
\delta / E_k = 1/2$ and Raman coupling strength $\hbar \Omega /
E_k$ varying between 0 and 2.  For this case of $\delta > 0$, a
condensate is formed in the ${\bf{Q}} \simeq - {\bf{k}}/2$ state
of the lower dispersion relation.  In the absence of Raman
coupling, this equality is exact, and the condensate is formed in
the $|B\rangle$ internal state. Quasiparticle excitations in the
$|B\rangle$ state follow the standard Bogoliubov result, while
quasiparticles in state $|A\rangle$ have a free-particle
dispersion relation -- this can be though of as due to the lack
of a spatial interference term between the condensate and
excitations in the $|A\rangle$ state at any wavevector.  For
small Rabi coupling strengths, one speaks more properly of
excitations in the lower and the upper quasiparticle dispersion
relations, although the energies and internal-state compositions
of these states remain relatively unchanged for wavevectors
${\bf{q}}$ far from the Doppler-shifted Raman resonance condition
$\delta - {\bf{q}} \cdot {\bf{k}}/m = 0$.  The lower quasiparticle
dispersion relation has a local minimum near ${\bf{q}} =
{\bf{k}}/2$ caused by the minimum of the dispersion relation for
the bare $|A\rangle$ state atoms.  By analogy to the roton
minimum in the excitation spectrum of superfluid $^4$He
\cite{nozi90}, we denote this feature as an ``artificial roton,''
although we mean to suggest no similarity between the specific
structure of a roton and of the plane-wave excitations of a
PDBEC.  As the Rabi frequency is increased further, a greater
mixing of the internal state composition of the Bose condensate
and of the quasiparticle excitations occurs, and energy levels
are shifted. In particular, the energy of the ``artificial
roton'' is shifted upwards due to interactions as the internal
state composition of the Bose condensate and of these excitations
becomes more similar.

\section{Impurity scattering}

We now consider the scattering of impurities passing through a
uniform PDBEC.  We consider impurities of mass $M_c$ which have a
free-particle dispersion relation $E_c = \hbar^2 k_c^2 / 2 M_c$
where $k_c$ is the impurity wavevector. Such impurities interact
with the two components of the periodically-dressed condensate
through an interaction of the form
\begin{equation}
{\mathcal{H}}_{\mbox{\small{scat}}} = \lambda \sum_{{\bf{m}},
{\bf{l}}, {\bf{q}}} \left( a^{\dagger}_{{\bf{m}} + {\bf{q}}}
a_{\bf{m}} c^{\dagger}_{{\bf{l}} + {\bf{q}}} c_{\bf{l}} +
b^{\dagger}_{{\bf{m}} + {\bf{q}}} b_{\bf{m}} c^{\dagger}_{{\bf{l}}
+ {\bf{q}}} c_{\bf{l}} \right) \label{eq:hscat}
\end{equation}
Here $a_{\bf{q}}$, $a_{\bf{q}}^\dagger$, $b_{\bf{q}}$, and
$b_{\bf{q}}^\dagger$ are defined as above, and  $c_{\bf{q}}$ and
$c^\dagger_{\bf{q}}$ are the creation and annihilation operators,
respectively, for the impurity particles.  The coupling strength
for the impurity scattering is $\lambda$, which we parameterize
as $\lambda = 2 \pi \hbar^2 a_c / M^\prime$.  This
parameterization correctly describes low-energy $s$-wave
scattering with a scattering length of $a_c$ and the reduced mass
$M^\prime = m M / (m + M)$.  Note that we have made the
assumption that the scattering strengths between the impurity and
atoms in states $A$ and $B$ are equal, and assume further that
state changing collisions do not occur (the impurity is considered
``non-magnetic'').  We may consider also the limit as the impurity
mass becomes infinite while the impurity velocity ${\bf{v}} =
\hbar {\bf{k}} / m$ remains constant (${\bf{k}}$ is the impurity
wavevector).  We thereby obtain a description of the PDBEC
flowing past a rigid obstacle with a velocity $\bf{v}$, and the
concept of impurity scattering is replaced by the dissipation of
superfluid flow.

The scattering rate of these impurities can be calculated using a
perturbative approach as follows \cite{timm97symp,kett00coll}.  We
consider the system to be initially in the state $ |0\rangle_{AB}
|{\bf k}\rangle_C$, i.e.\ a PDBEC with no excitations, and a
single impurity particle of wavevector ${\bf{k}}$. The scattering
Hamiltonian of Eq. \ref{eq:hscat} may couple this state to the
final state $|f\rangle_{AB} |{\bf{k}} - {\bf{q}} \rangle_C$ in
which the periodically-dressed gas is in the state $|f\rangle$
and the impurity wavevector changes from ${\bf{k}}$ to
${\bf{k}}-{\bf{q}}$. The matrix element for this transition is
$\lambda \langle f | n_{\bf{q}}|0\rangle$, where $n_{\bf{q}}$ is
described above, and where  the $AB$ subscript labeling the
quantum states of the periodically-dressed gas has been dropped.
We then obtain the rate of impurity scattering by summing over
all final states $|f\rangle$ and momentum transfers $\bf{q}$
using Fermi's Golden Rule.

We now make use of the Bogoliubov approximation. If we consider
only weak scattering, we may restrict the final states to those
containing a single quasiparticle, i.e.\ $|f\rangle =
|{\bf{q}}^\pm\rangle$ where $|{\bf{q}}^- \rangle =
\tilde{\mu}_{{\bf{Q}}+\bf{q}}^\dagger |0 \rangle$ and $|{\bf{q}}^+
\rangle = \tilde{\pi}_{{\bf{Q}}+\bf{q}}^\dagger |0 \rangle$.  The
rate of impurity scattering is then
\begin{equation}
\Gamma = \frac{2 \pi}{\hbar^2} \lambda^2 \sum_{\bf{q}} \left[
S_\mu({\bf{q}}) \delta \left(\tilde{\omega}_{{\bf{Q}}+\bf{q}}^- -
{\bf{v}} \cdot {\bf{q}} + \frac{\hbar q^2}{2 M_C} \right) +
S_\pi({\bf{q}}) \delta \left(\tilde{\omega}_{{\bf{Q}}+\bf{q}}^+ -
{\bf{v}} \cdot {\bf{q}} + \frac{\hbar q^2}{2 M_C} \right) \right]
\label{eq:gammaeq}
\end{equation}
where ${\bf{v}} = \hbar {\bf{k}}/M$ is the impurity velocity.  We
have introduced the quantities $S_{\mu}({\bf{q}})$ and
$S_\pi({\bf{q}})$ which relate to the static structure factor
describing density-density correlations in the Bose condensate.
For a single-component Bose-Einstein condensate, the static
structure factor $S(\bf{q})$ has been evaluated and measured
\cite{stam99phon,zamb00struc,stei02exc}.  For the PDBEC, we
separate the contributions of the lower and upper excitation
spectra as
\begin{eqnarray}
S_\mu({\bf{q}}) & = & \frac{1}{N_0} \langle 0 | n_{-\bf{q}} |
{\bf{q}}^- \rangle \langle {\bf{q}}^- | n_{\bf{q}} | 0 \rangle \\
S_\pi({\bf{q}}) & = & \frac{1}{N_0} \langle 0 | n_{-\bf{q}} |
{\bf{q}}^+ \rangle \langle {\bf{q}}^+ | n_{\bf{q}} | 0 \rangle
\end{eqnarray}
Physically, these quantities describe two effects.  First, the
magnitude of $S_\mu$ and $S_\pi$ describes the contributions of
quasiparticles in the lower ($\mu$) or upper ($\pi$)
quasiparticle state to density fluctuations (as opposed to phase
fluctuations, which dominate, for example, for phonon excitations
of a scalar Bose condensate
\cite{stam99phon,stei02exc,dett01phase}). Second, these quantities
describe the degree to which the internal state composition of
the condensate matches that of the quasiparticle.   For instance,
a condensate which is predominantly in the $|B\rangle$ internal
state will not scatter strongly into quasiparticle states in
which the $|A\rangle$ internal state is dominant, and thus the
structure factor describing this scattering will be small.

Before proceeding to a calculation of $S_\mu$ and $S_\pi$, it is
useful to consider the scattering rate $\Gamma_0$ for an
ideal-gas, single-component Bose-Einstein condensate as the
target gas.  In this case, we may conveniently take
$S_\mu({\bf{q}}) = 1$, $S_\pi({\bf{q}}) = 0$ and
$\tilde{\omega}^-_{\bf{q}} = \hbar q^2 / 2 m$. Replacing
$\sum_{\bf{q}} = V/(2 \pi)^3 \int d^3 {\bf{q}}$, and substituting
for $\lambda$ we evaluate Eq.\ \ref{eq:gammaeq} and obtain simply
\begin{equation}
\Gamma_0 = \frac{\lambda^2 N_0 V {M^\prime}^2 v}{\pi \hbar^4} =
\frac{N_0}{V} \times 4 \pi a_c^2 \times v,
\end{equation}
i.e.\ the conventional expression for the scattering rate where
$N_0/V$ is the density of the target gas and $4 \pi a_c^2$ is the
collision cross section.  We may then determine the impurity
scattering rate in a PDBEC relative to that in an ideal gas,
$F_{\mbox{\small tot}} = \Gamma/\Gamma_0 = F_\mu + F_\pi$:
\begin{eqnarray}
F_\mu & = & \frac{\hbar^2}{4 \pi v {M^\prime}^2} \int d^3 {\bf{q}}
S_\mu({\bf{q}}) \delta\left( \tilde{\omega}_{{\bf{Q}}+\bf{q}}^- +
\frac{\hbar q^2}{2 M_C} - {\bf{v}} \cdot {\bf{q}} \right)
\label{eq:fmuexpression}
\\
F_\pi & = & \frac{\hbar^2}{4 \pi v {M^\prime}^2} \int d^3
{\bf{q}} S_\pi({\bf{q}}) \delta\left(
\tilde{\omega}_{{\bf{Q}}+\bf{q}}^+ + \frac{\hbar q^2}{2 M_C} -
{\bf{v}} \cdot {\bf{q}} \right) \label{eq:fpiexpression}
\end{eqnarray}

\section{Structure factor formalism}

To calculate the structure factors $S_\mu$ and $S_\pi$ we start
by expressing the density operator $n_{\bf{q}}$ in terms of the
quasiparticle operators
\begin{eqnarray}
n_{\bf{q}} & = & \frac{1}{2} \sum_{\bf{m}} \left( \begin{array}
{c c c c} a^\dagger_{{\bf{Q}} + {\bf{m}} + {\bf{q}}} & -i
a_{{\bf{Q}} - {\bf{m}} - {\bf{q}}} & b^\dagger_{{\bf{Q}} +
{\bf{m}} + {\bf{q}}}  & -i b_{{\bf{Q}} - {\bf{m}} - {\bf{q}}}
\end{array} \right) \left( \begin{array}{c}
a_{{\bf{Q}} + {\bf{m}} } \\
i a^\dagger_{{\bf{Q}} - {\bf{m}}} \\
b_{{\bf{Q}} + {\bf{m}} } \\
i b^\dagger_{{\bf{Q}} - {\bf{m}}}
\end{array} \right) \\
& = & \frac{1}{2} \sum_{\bf{m}}  \left( \begin{array} {c c c c}
\tilde{\mu}^\dagger_{{\bf{Q}} + {\bf{m}} + {\bf{q}}} & -i
\tilde{\mu}_{{\bf{Q}} - {\bf{m}} - {\bf{q}}} &
\tilde{\pi}^\dagger_{{\bf{Q}} + {\bf{m}} + {\bf{q}}}  & -i
\tilde{\pi}_{{\bf{Q}} - {\bf{m}} - {\bf{q}}}
\end{array} \right) M_{{\bf{q}} + {\bf{m}}}^\dagger R_{{\bf{q}} + {\bf{m}}}^\dagger R_{\bf{m}} M_{\bf{m}} \left( \begin{array}{c}
\tilde{\mu}_{{\bf{Q}} + {\bf{m}} } \\
i \tilde{\mu}^\dagger_{{\bf{Q}} - {\bf{m}} } \\
\tilde{\pi}_{{\bf{Q}} + {\bf{m}} } \\
i \tilde{\pi}^\dagger_{{\bf{Q}} - {\bf{m}} }
\end{array} \right)
\end{eqnarray}
Acting on a PDBEC, the only important terms in the sum are those
which include the condensate operators $\tilde{\mu}_{\bf{Q}}$ and
$\tilde{\mu}_{\bf{Q}}^\dagger$.  Applying the relations
$\tilde{\mu}_{{\bf{Q}}+{\bf{q}}} |0\rangle =
\tilde{\pi}_{{\bf{Q}}+{\bf{q}}} |0\rangle = 0$ for ${\bf{q}} \neq
0$, and the relation  $M_{\bf{0}} = \bf{1}$, we define matrices
$\alpha = M_{\bf{q}}^\dagger R_{\bf{q}}^\dagger R_{\bf{0}}$ and
$\beta = R_{\bf{0}}^\dagger R_{-\bf{q}} M_{-{\bf{q}}}$ and obtain
\begin{equation}
n_{\bf{q}} |0\rangle = \frac{\sqrt{N_0}}{2} \left( (\alpha_{1 1}
+ i \alpha_{1 2} + i \beta_{1 2} + \beta_{2 2})
\tilde{\mu}^\dagger_{\bf{q}}|0\rangle + (\alpha_{3 1} + i
\alpha_{3 2} + i \beta_{1 4} + \beta_{2 4})
\tilde{\pi}^\dagger_{\bf{q}}|0\rangle \right)
\end{equation}
So
\begin{eqnarray}
S_\mu({\bf{q}}) & = & \frac{1}{4} \left|\alpha_{1 1} + i
\alpha_{1 2} + i \beta_{1 2} + \beta_{2 2}\right|^2 \\
S_\pi({\bf{q}}) & = & \frac{1}{4} \left|\alpha_{3 1} + i
\alpha_{3 2} + i \beta_{1 4} + \beta_{2 4}\right|^2
\end{eqnarray}

\begin{figure}[htbf]
\begin{center}
\includegraphics[width=5in]{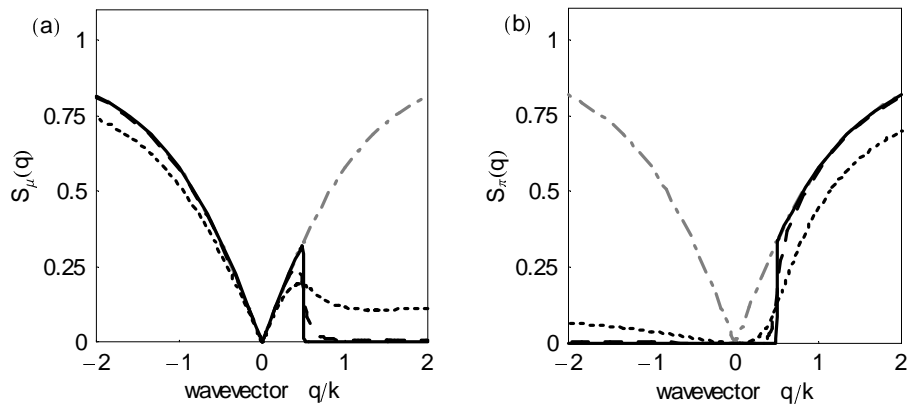}
\end{center}
     \caption{Variations of the static structure factors (a) $S_\mu({\bf{q}})$ and (b) $S_\pi({\bf{q}})$
     with the Raman coupling strength $\Omega$.
     The wavevector $\bf{q}$ is colinear with the Raman momentum transfer ${\bf{k}}$ with $q={\bf{q}} \cdot {\bf{k}}$.
     Three black curves describe a PDBEC with $\hbar \delta/E_k = 1/2$, $\mu/E_k = 1$,
     and $\hbar \Omega /E_k = 0$ (solid), $1/4$ (dashed) and $1$ (dotted).  Scattering for $q \sim 0$ and $q < 0$ occurs
     only into the lower
     quasiparticle branch and for $q \gg k$ into the upper quasiparticle branch.  In other
     regions, stronger Raman coupling leads to eigenstates with mixed populations in the internal states $|A\rangle$ and
     $|B\rangle$ over a broader range of $q$, and thus impurity scattering occurs into both the lower and upper quasiparticle
     branches.  Also shown is the structure factor $S({\bf{q}})$
     for a scalar
     Bose-Einstein condensate (gray dot-dashed curve) with $\mu/E_k = 1$.
     }
     \label{fig:smuandspivaryrabi}
\end{figure}

\begin{figure}[htbf]
\begin{center}
\includegraphics[width=5in]{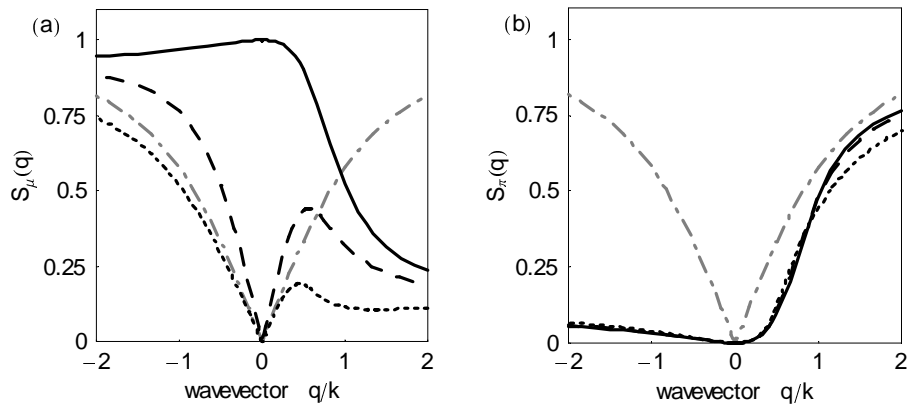}
\end{center}
     \caption{Static structure factor (a) $S_\mu({\bf{q}})$ and (b) $S_\pi({\bf{q}})$
     for a PDBEC.  The wavevector $\bf{q}$ is colinear with the Raman momentum transfer ${\bf{k}}$ with $q={\bf{q}} \cdot {\bf{k}}$.
     Three black curves describe a PDBEC with $\hbar \delta/E_k = 1/2$, $\hbar \Omega/E_k = 1$,
     and $\mu/E_k = 0$ (solid), $1/4$ (dashed) and $1$ (dotted).  The condensate in the $\mu_{\bf{Q}}$ state
     is predominantly in the internal state $B$; thus scattering predominantly occurs into the lower
     quasiparticle branch for $q < 0$ and into the upper quasi-particle branch for large $q>0$.  For
     $q \sim 0.5$
     both $S_\mu({\bf{q}})$ and $S_\pi({\bf{q}})$ are significant since quasiparticles contain large fractions
     of both $|A\rangle$ and $|B\rangle$.
     Interactions suppress scattering at small momentum transfer, an effect seen also in the structure factor $S({\bf{q}})$
     for a scalar
     Bose-Einstein condensate (gray dot-dashed curve) with $\mu/E_k = 1$.
     }
     \label{fig:smuandspivarymu}
\end{figure}

The static structure factors $S_\mu({\bf{q}})$ and
$S_\pi({\bf{q}})$ for a PDBEC are shown in Figures
\ref{fig:smuandspivaryrabi} and \ref{fig:smuandspivarymu}.  The
specific values for the Raman detuning $\delta$, the Rabi
frequency $\Omega$ and the chemical potential $\mu$ are relevant
for a particular realization of a PDBEC using two hyperfine
states of a $^{87}$Rb, as discussed above.  The division of the
scattering weight between the lower ($S_\mu$) and upper ($S_\pi$)
quasiparticle branches can be understood by considering the
composition of the dressed-state quasiparticles, which is related
by Eq.\ \ref{eq:rotation} to the mixing angle
$\theta_{\bf{q}}$.   This angle ranges from $\theta_{\bf{q}} = 0$
for ${\bf{q}} \cdot {\bf{k}} \rightarrow - \infty$ to
$\theta_{\bf{q}} = \pi$ for ${\bf{q}} \cdot {\bf{k}} \rightarrow
+ \infty$, reaching $\theta_{\bf{q}} = \pi/2$ at the
Doppler-shifted Raman resonance $\delta - {\bf{q}} \cdot
{\bf{k}}/m = 0$.  For small Rabi frequency, $\theta_{\bf{q}}$,
and hence the internal state composition of the dressed states
varies rapidly about the Raman resonance condition, while for
large Rabi frequency, this variation occurs over a broader range
of $q$.  This behaviour is reflected in the structure factors, as
illustrated by Figure \ref{fig:smuandspivaryrabi}.  By
continuity, the internal state composition of the condensate in
the $\mu_{\bf{Q}}$ state is nearly identical to that of
excitations in the lower quasiparticle branch (created by
$\tilde{\mu}_{\bf{Q} + \bf{q}}^\dagger$), and nearly orthogonal
to that of excitations in the upper quasiparticle branch (created
by $\tilde{\pi}_{\bf{Q} + \bf{q}}^\dagger$).  Thus, since elastic
scattering from a ``non-magnetic'' impurity does not affect the
internal state distribution,  $S_\pi({\bf{q}}) \simeq 0$
generally for small $q$.  Away from $q = 0$, the division of the
scattering weight between the upper and lower quasiparticle
branches depends on the strength of the Rabi coupling.  For small
Rabi frequencies, the scattering strength shifts abruptly from the
lower to the upper quasiparticle branch at the Doppler-shifted
Raman resonance.  For large Rabi frequency, the changes in the
scattering strengths are more gradual, with scattering into both
branches showing a significant weight for a wide range about the
Raman resonance condition.  This variation is particularly
relevant for the scattering of impurities near the Landau
critical velocity, as discussed below.

As illustrated in Figure \ref{fig:smuandspivarymu}, interatomic
interactions suppress scattering at small momentum transfers. For
a single-component homogeneous Bose-Einstein condensate, the
structure factor $S({\bf{q}}) = q^2 / \sqrt{q^2 ( q^2 + 2
\xi^{-2})}$ is small for wavevectors of magnitude $q < 1/\xi$
where $\xi = \sqrt{2 m \mu / \hbar^2}$ is the healing length.
Similarly, for the PDBEC, we find a suppression of the structure
factors $S_\mu({\bf{q}})$ and $S_\pi({\bf{q}})$ for $q < k
\sqrt{2\mu/E_k}$, as shown in the figure.  This suppression of
the structure factor in a PDBEC could be probed experimentally as
it was for a scalar Bose-Einstein condensate by studying
shallow-angle optical Bragg scattering
\cite{stam99phon,stei02exc}.

\section{Dissipation of superfluid flow}

Finally, let us use the formalism developed above to describe
superfluid properties of a PDBEC.  In Ref.\ \cite{higb02}, we
applied the Landau criterion to determine the critical velocity
for superfluidity in a PDBEC given the quasiparticle dispersion
relations.  This criterion gives the superfluid critical velocity
along a direction $\hat{{\bf{q}}}$ as having a magnitude $v_c =
\min_{q,i} E_i(q {\hat{\bf{q}}}) / \hbar q$ where $E_i({\bf{q}})$
are the energies of possible excitations with wavevector
${\bf{q}}$; using the momentum definitions for a PDBEC, these
would be the quasiparticle energies $\hbar
\omega_{{\bf{Q}}+{\bf{q}}}^\pm$. In an unperturbed, single
component weakly-interacting condensate, this critical velocity
$v_c^0$ is equal to the Bogoliubov speed of sound $c_s =
\sqrt{\mu/m}$ in all directions. In a PDBEC, the anisotropic
quasiparticle dispersion relation leads directly to the
prediction of an anisotropic superfluid critical velocity, and to
the possibility of a critical velocity lower than that of the
unperturbed condensate.

One may determine this critical velocity graphically by finding
the line of minimum absolute slope which connects the origin and
the quasiparticle dispersion curve (see Figure
\ref{fig:quasidisprelations}).  For the PDBEC, the superfluid
critical velocity may be suppressed by the presence of the
``artificial roton'' feature in the lower dispersion curve. This
suppression occurs in the direction of the Raman momentum
transfer $\hbar {\bf{k}}$ for the case of $\delta > 0$ (as shown
in the figure), or in the opposite direction for $\delta < 0$.

However, while the Landau criterion determines the minimum
velocity for the dissipation of superfluid flow, it does not
quantify the onset of such dissipation as the velocity is
increased beyond the critical value.  Indeed, for extremely weak
Raman coupling, even though the superfluid critical velocity is
strictly different,  one should expect the fluid properties of a
PDBEC to be nearly identical to that of a two-component BEC in
the complete absence of Raman coupling.  We can quantitatively
assess this description using the formalism developed above for
the determination of the matrix elements $S_\mu$ and $S_\mu$, and
therefrom the normalized scattering rates $F_\mu$ and $F_\pi$.

\begin{figure}[htbf]
\begin{center}
\includegraphics[width=3in]{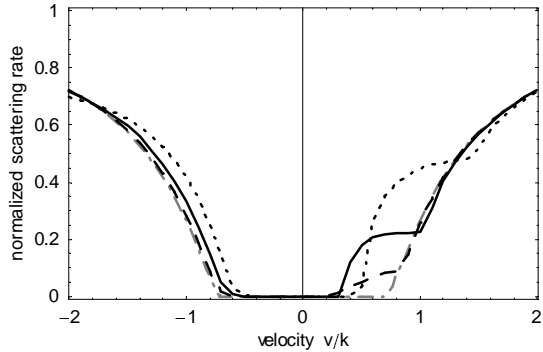}
\end{center}
     \caption{Dissipation strength of superfluid flow in periodically-dressed
     Bose-Einstein condensates.  The scattering rate $F_{tot}$ for infinitely massive impurities traveling
     through the condensate, normalized to the scattering in an ideal gas Bose-condensate at similar
     densities, is plotted vs.\ the velocity of the impurity.
     Here we consider velocities parallel to the Raman momentum transfer, with $v = {\bf{v}} \cdot {\bf{k}} / k$.
     Black curves show results for PDBEC's with varying strengths of the Raman coupling; for all curves, $\delta/E_k =
     1/2$ and
     $\mu/E_k = 1$ while $\hbar \omega / E_k = 1/2$ (dashed), 1 (solid) or 2 (dotted).  The gray dot-dashed curve shows results
     for a one-component condensate with $\mu/E_k = 1$.  The superfluid critical velocity of the PDBEC in the direction of the Raman
     momentum transfer is suppressed with respect to an uncoupled condensate.  For weakly Raman coupling, the dissipation
     at low velocities is weak.  This dissipation grows stronger as the Raman coupling strength is increased.
     }
     \label{fig:allpdplot}
\end{figure}

\begin{figure}[htbf]
\begin{center}
\includegraphics[width=3in]{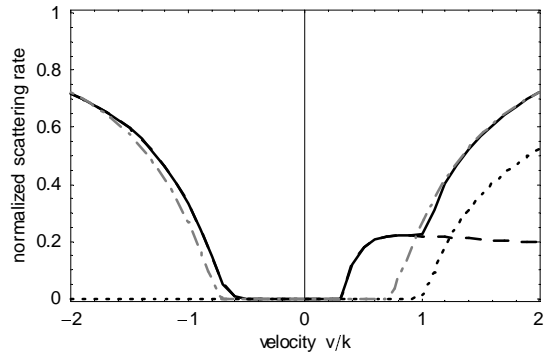}
\end{center}
     \caption{Normalized rates for production of lower ($F_\mu$, dashed line) and upper
     ($F_\pi$, dotted line) quasiparticle excitations due to scattering off infinitely massive impurities
     moving at velocities $v$ in a PDBEC.  Velocities parallel to the Raman momentum transfer are considered, with
     $v = {\bf{v}} \cdot {\bf{k}} / k$.  Black curves show results for a PDBEC with $\hbar \delta / E_k = 1/2$,
     $\mu/E_k = 1$, and $\hbar \Omega / E_k = 1$, with the solid curve showing the total normalized scattering
     rate $F_{\mbox{tot}}$.  The gray dot-dashed curve shows the normalized scattering rate for a one-component condensate
     with $\mu/E_k = 1$.  For $v>0$ (for this case where $\delta >0$), impurity scattering at low
     velocities is enhanced as compared to the one-component condensate.  This
     scattering leads to the creation of excitations in the lower
     dispersion curve ($F_\mu > 0$) around the ``artificial
     roton'' feature in the quasiparticle dispersion relations.  At velocities near the critical velocity for the
     unperturbed condensate, $v_c^0$, scattering is allowed into the upper dispersion relation $F_\pi > 0$, and the normalized
     scattering rate approaches that of a one-component condensate.  For $v<0$, energy and momentum conservation
     imply that only scattering into the lower quasiparticle excitations is allowed at the velocities considered.
     }
     \label{fig:onepdplot}
\end{figure}

The results of such calculations for infinitely massive
impurities (i.e.\ superfluid flow about rigid obstacles) are
presented in Figures \ref{fig:allpdplot} and
\ref{fig:onepdplot}.  Values of $F_\mu$ and $F_\pi$ were
determined by numerical integration in which the delta function
in Eqs. \ref{eq:fmuexpression} and \ref{eq:fpiexpression} was
approximated by a narrow Gaussian distribution whose width was
adjusted to give reasonable numerical convergence.  As we expect,
for small Rabi frequencies $\Omega$ the properties of a
periodically-dressed BEC are quite similar to those of an
uncoupled two component Bose-Einstein condensate. The scattering
probability for impurity atoms (related to the viscosity of the
fluid) is indeed non-zero at velocities lower than the superfluid
critical velocity for the unperturbed gas, demonstrating that the
critical velocity is suppressed along the direction of Raman
scattering. This scattering at low velocities occurs through the
excitation of quasiparticles near the local minimum (the
``artificial roton'') in the lower dispersion relation.  However,
this scattering channel is weak for small values of the Rabi
frequency $\Omega$, and thus the scattering strength
$F_{\mbox{tot}}$ remains weak until the curve joins that of the
uncoupled condensate and the scattering strength is increased by
phonon scattering.

The effects of increasing the Rabi frequency are twofold. First,
as $\Omega$ increases, the internal state composition of the
``artificial roton'' excitation becomes more similar to the
internal state composition of the Bose-Einstein condensate.  This
causes an increase in the scattering strength  (as indicated also
in the calculations of $S_\mu$ shown in Figures
\ref{fig:smuandspivarymu} and \ref{fig:smuandspivaryrabi}).
Simultaneously, because of this overlap of the internal state
composition, the energy of these excitations is increased,
approaching the energy given by the one-component Bogoliubov
spectrum, albeit with an appropriately modified effective mass.
Thus, the minimum velocity at which scattering occurs increases
as the Rabi frequency increases.  At moderate values of the Rabi
frequency (where $\hbar \Omega/E_k \sim 1$), the scattering rate
vs.\ velocity has a two-step shape as scattering is first
enhanced by strong scattering into the ``artificial roton''
state, and then later enhanced by phonon scattering when the
velocity is near $v_c^0$.  As shown in Figure \ref{fig:onepdplot}
for one typical parameter setting, only $F_\mu$ contributes to
the weak scattering at velocities smaller than $v_c^0$.  At
velocities greater than $v_c^0$, scattering into the upper
dispersion curve is allowed and $F_\pi \neq 0$.

This two-step curve gives a clear signature which should be
visible in experimental probes of superfluid flow.  Indeed,
experiments such as those performed by Chikkatur {\emph et al}
\cite{chik00}, in which atoms of another ground state hyperfine
level were used as scattering impurities in a dilute atomic Bose
condensate,  could be carried out in a straightforward manner.  It
would be interesting to consider further how other aspects of
superfluidity, such as the onset of turbulent flow \cite{rama99}
or the metastability of quantized vortices
\cite{matt99vort,madi99vort}, are affected by the anisotropic
nature of a periodically-dressed Bose-Einstein condensate.

In conclusion, dissipation of superfluid motion in a
periodically-dressed Bose-Einstein condensate above the
superfluid critical velocity was treated.  A formalism was
developed for the calculation of the static structure factors
$S_{\mu}$ and $S_{\pi}$ describing density-density correlations
from quasiparticles in the lower ($\mu$) and upper ($\pi$)
excitation branches of the periodically coupled gas.  Following a
perturbation theory approach, the scattering rate of massive
impurities, or equivalently the damping rate of superfluid motion
past microscopic obstacles, was calculated.  Raman coupling of a
two-component condensed gas leads to a suppression of the
superfluid critical velocity.  The impurity scattering rate
evolves smoothly as the Raman coupling strength is increased. The
the superfluid damping is weak for weak Raman coupling, and
becomes more significant as the Raman coupling strength is
increased.

\ack{I thank James Higbie for insightful discussions. Support
from the NSF, the Sloan Foundation, the Packard Foundation, and
the University of California Hellman Family Faculty Fund is
gratefully acknowledged.}
\bibliographystyle{prsty}

\begin{thebibliography}{10}
\bibitem{degeneracyfootnote}
We do not consider here the possibility of a degenerate or
near-degenerate
  ground state. Proper treatment of a periodically-dressed Bose-Einstein
  condensate for such a situation requires a many-body treatment beyond the
  Bogoliubov approximation, as will be discussed elsewhere.
\end{thebibliography}


\end{document}